\documentclass[10pt,aps,prc,twocolumn,showpacs,superscriptaddress,floatfix,nofootinbib]{revtex4-1}   

\usepackage{graphicx}  
\usepackage{dcolumn}   
\usepackage{bm}        
\usepackage{amssymb}   
\usepackage{amsmath}
\usepackage{amsfonts}
\usepackage{graphics}
\usepackage{epsfig}
\usepackage[normalem]{ulem}
\usepackage[usenames]{color}
\usepackage{multirow}
\usepackage{hyperref}
\usepackage{color}  
\usepackage{lipsum}
\usepackage{lineno}
\usepackage{verbatim}
\usepackage{subfigure}
\usepackage[noend]{algorithmic}
\usepackage[noend]{algorithm2e}

\def\pT{\mbox{$p_{\rm T}$}}

\def\KTTwo{\mbox{$K_{\rm T2}$}}

\def\vecp{\mbox{${\bf p}$}}
\def\<{\mbox{$\langle$}}
\def\>{\mbox{$\rangle$}}
\def\o{\mbox{$\omega$}}
\def\a{\mbox{$\alpha$}}
\def\b{\mbox{$\beta$}}
\def\g{\mbox{$\gamma$}}
\def\t{\mbox{$\tau$}}
\def\e{\mbox{$\epsilon$}}
\def\dag{\mbox{$\dagger$}}

\DeclareMathSymbol{\mlq}{\mathord}{operators}{``}
\DeclareMathSymbol{\mrq}{\mathord}{operators}{`'}

\makeatletter

\newcommand{\Rmnum}[1]{\expandafter\@slowromancap\romannumeral #1@}
\makeatother

\begin{document}          

\title{Multipion correlations induced by isospin conservation of coherent emission}
\author{Dhevan Gangadharan}
\email[]{dhevanga@gmail.com}
\affiliation{Nuclear Science Division, Lawrence Berkeley National Laboratory, 1 Cyclotron Rd, Berkeley, CA 97420}

\begin{abstract}
Recent measurements have revealed a significant suppression of multipion Bose-Einstein correlations in heavy-ion collisions at the LHC.  
The suppression may be explained by postulating coherent pion emission.
Typically, the suppression of Bose-Einstein correlations due to coherence is taken into account with the coherent state formalism in quantum optics.  
However, since charged pion correlations are most often measured, the additional constraint of isospin conservation, which is absent in quantum optics, needs to be taken into account.
As a consequence, correlations emerge between pions of opposite charge.
A calculation of the correlations induced by isospin conservation of coherent emission is made for two, three- and four-pion correlation functions and compared to the data from the LHC.

\end{abstract}
\maketitle

\section{Introduction}
Data from high energy collisions present a unique opportunity to study the possibility of quantum coherence at very high temperatures.
Bose-Einstein (BE) correlations of identical bosons are often used to search for coherent emission.
In the presence of coherent emission, the strength of BE correlations is expected to be suppressed.
A $6\sigma$ suppression of four-pion BE correlations was recently found in Pb--Pb collisions at the LHC with ALICE \cite{Adam:2015pbc}.

Multibody Coulomb correlations have also been proposed as a potential source of the suppression. 
However, such correlations are unlikely to decrease with $K_{\rm T}$, in contrast with the observed decrease of the suppression at high $K_{\rm T}$.
An important additional feature of BE correlations in high multiplicity collisions is their robustness to background correlations unrelated to quantum statistics (QS) and final-state interactions (FSI).
The QS and FSI correlations for large emission volumes occur in a very narrow interval in relative momentum while all known background correlations produce easily distinguished broad correlations.

While two-pion BE correlations are most often measured experimentally, they are insufficient to search for coherent emission.  
The unknown freeze-out distribution of particles produced in collisions causes two noteworthy uncertainties in a two-pion analysis.
The first being the fraction of pions from short- compared to long-lived emitters, which characterizes a dilution of BE correlations.
Secondly, the computation of FSI is done with an assumed freeze-out distribution.  
Both of these uncertainties make it practically impossible to determine the presence of coherent emission from two-pion correlations alone.
However, both uncertainties were found to largely cancel in the comparison of measured to {\it built} multipion BE correlations \cite{Adam:2015pbc, Gangadharan:2015ina}.

The effect of pion coherence on BE correlations is typically incorporated using the coherent state formalism of quantum optics \cite{Glauber:1963fi}.
However, the fact that {\it charged} pion correlations are measured necessitates an extension of the quantum optics approach using the super-selection rule \cite{Horn:1971wd, Botke:1974ra, Gyulassy:1979yi, Biyajima:1998yh, Akkelin:2001nd}.
The resulting correlations induced by isospin conservation of the coherent component occurs between all pion species.
Isospin conservation of the chaotic component also induces additional correlations in scenarios where the emission duration of the source is short \cite{Andreev:1991xf}.
In this letter, expressions are derived for three- and four-pion correlations stemming from isospin conservation of pion coherent states.
Calculations are presented for four different mixed-charge correlation functions and compared to the LHC data.

\section{Formalism}
The formalism of pion coherent states obeying the super-selection rule and their application to BE correlations is given in detail in Refs.~\cite{Biyajima:1998yh} and \cite{Akkelin:2001nd}.
It is assumed that particle production can be split into a chaotic and a single coherent component, whose annihilation operators at a given momentum are given by $b(p)$ and $d(p)$, respectively.  
\begin{eqnarray}
a_i(p) = b_i(p) + e_i d(p) \\
e_0 = \cos(\theta) \\
e_{\pm} = \frac{\sin{\theta}}{\sqrt{2}} e^{\pm \phi}.
\end{eqnarray}
The subscript $i$ denotes the pion type ($\pi^+,\pi^-,\pi^0$) for which one commonly introduces a unit vector, ${\bf e}$, in three-dimensional isospin space.
The single particle inclusive momentum densities are given in the usual way 
\begin{eqnarray}
N_i^{(1)}(p) \equiv E_{\vecp}\frac{d^3N_i}{d^3\vecp} = \< a_i^{\dag}(p) a_i(p) \>, \nonumber \\
= \< b_i^{\dag}(p) b_i(p) \> + \<|e_i|^2 \> \< d^{\dag}(p) d(p) \>, \nonumber \\
= N_{\rm ch}^{(1)}(p) + N_{\rm coh}^{(1)}(p),
\end{eqnarray}
where $E_{\vecp} = \sqrt{ m^2 + \vecp^2}$.  
The chaotic and coherent components are given by $\rm ch$ and $\rm coh$, respectively.
An averaging over all possible orientations of the isospin vector is done in order to compute all final observables.
The following integrals over the isospin vector will be needed to evaluate the two-, three-, and four-pion correlation functions.
\begin{eqnarray}
\< |e_+|^n \> &=& \frac{1}{4\pi} \int d\cos(\theta)d\phi \left[ \frac{\sin{\theta}}{\sqrt{2}} \right]^n \\
\< |e_+|^2 \> &=& \< |e_-|^2 \> = \< |e_0|^2 \> = \frac{1}{3} \\
\< |e_+|^4 \> &=& \frac{2}{15} \quad \< |e_+|^6 \> = \frac{2}{35} \quad \< |e_+|^8 \> = \frac{8}{315} \label{eq:en}
\end{eqnarray}
Integrals which contain a mixture of $e_+$ and $e_-$ are identical to the ones given above.
The total number of pions which are radiated from the classical source at momentum \vecp~is given by $|d(\vecp)|^2$ while the coherent fraction of pions is defined as $G(p) \equiv \frac{N_{\rm coh}^{(1)}(p)}{N^{(1)}(p)} = \frac{1}{3} |d(\vecp)|^2$.

It is convenient to introduce the single particle Wigner function, split into chaotic and coherent components
\begin{equation}
f_{{\bf e},i}(x,p) = f_{\rm ch}(x,p) + |e_i|^2f_{\rm coh}(x,p),
\end{equation}
which provide the following two important links between the expectation values of the pionic field operators and an integration over the freeze-out hypersurface ($\sigma_{\rm out}$),
{\footnotesize
\begin{eqnarray}
\< b_i^{\dag}(p_1) b_i(p_2) \> = \int_{\sigma_{\rm out}} d\sigma_{\mu} p^{\mu} f_{\rm ch}(x,p) e^{-iqx} \nonumber \\ 
\equiv T_{12}e^{-i\Phi_{12}} \sqrt{[1-G(p_1)][1-G(p_2)]N_i^{(1)}(p_1)N_i^{(1)}(p_2)}, \label{eq:Tij} \\
\< d_i^{\dag}(p_1) d_i(p_2) \> = \< |e_i|^2 \> \int_{\sigma_{\rm out}} d\sigma_{\mu} p^{\mu} f_{\rm coh}(x,p) e^{-iqx} \nonumber \\
\equiv t_{12}e^{-i\phi_{12}} \sqrt{G(p_1)G(p_2)N_i^{(1)}(p_1)N_i^{(1)}(p_2)} \label{eq:tij},
\end{eqnarray}}
where $q=p_1-p_2$ and $p=(p_1+p_2)/2$.
The pair exchange magnitudes of the chaotic and coherent components are denoted by $T_{ij}$ and $t_{ij}$, respectively.
For the expectation value of two or more coherent pions with an imbalance of operators at momentum $p_1$ and $p_2$, we have the relation
{\footnotesize
\begin{eqnarray}
\< d_{\o_1}^{\dag}(p_1) d_{\o_1}(p_2) \cdot d_{\o_2}^{\dag}(p_3) d_{\o_2}(p_3) \cdot ... \cdot d_{\o_n}^{\dag}(p_n) d_{\o_n}(p_n) \> \nonumber \\
= \< \prod_\gamma^n | e_{\o_\gamma} |^2 \> \left[ \int  d\sigma_{\mu}p^{\mu}f_{\rm coh}(x,p) e^{-iq_{12}x} \right] \< d^{\dag}(p) d(p) \>^{n-1} \label{eq:tijmixed}
\end{eqnarray}}

\section{Three- and four-pion quantum statistics correlation functions}
The multipion inclusive momentum density distributions is given in the usual way as 
\begin{eqnarray}
N_{\o_1...\o_n}^{(n)}(p_1,...,p_n) = \left[ \prod_{\alpha=1}^{n} E_{\vecp_{\alpha}} \right] \frac{d^{3n}N_{\o_1...\o_n}}{\prod_{\alpha=1}^{n} d^3\vecp_{\alpha}} \nonumber \\
= \< \prod_{\alpha=1}^{n} a_{\o_{\alpha}}^{\dag}(p_{\alpha}) a_{\o_{\alpha}}(p_{\alpha}) \>,
\end{eqnarray}
where $\o$ represents the set of $n$ elements, each of which denote a particular type of pion.  
For example, the set $\o$ in the case of the $\pi^+ \pi^+ \pi^-$ distribution is given by $\o_1=\pi^+, \o_2=\pi^+, \o_3=\pi^-$.

Experimentally, the multipion distributions are often projected onto the Lorentz invariant relative momentum and average pair transverse momentum defined by 
\begin{eqnarray}
Q_n = \sqrt{- \sum_{i=1}^{n-1} \sum_{j=i+1}^{n} (p_{i} - p_{j})^2}, \\
K_{\rm Tn} = |\sum_{i=1}^{n} \vecp_{{\rm T},i}| / n.
\end{eqnarray}
The three- and four-pion QS distributions are decomposed into several components,
\begin{eqnarray}
N_{ijk}^{(3)} \equiv I_1 + I_2 + I_3, \\
N_{ijkl}^{(4)} \equiv J_1 + J_2 + J_3 + J_4, \label{eq:J1234}
\end{eqnarray}
where the $I_1$ and $J_1$ will be defined to contain the conventional QS correlations as prescribed by quantum optics.
The other components arise from the constraint of isospin conservation of coherent emission.
Neglecting FSI, the components for three-pion correlations are given in Eqs.~\ref{eq:I1}-\ref{eq:I3} while those for four-pion correlations are given in the appendix.
{\scriptsize
  \begin{widetext}
    \begin{eqnarray}
      I_1 &=& N_i^{(1)}(p_1)N_j^{(1)}(p_2)N_k^{(1)}(p_3) + \sum_{\o} \delta_{\o_{\alpha}\o_{\beta}} \Big[ |\< b_{\o_{\alpha}}^{\dag}(p_{\alpha}) b_{\o_{\alpha}}(p_{\beta}) \>|^2 + 2\Re \Big\{ \< b_{\o_{\alpha}}^{\dag}(p_{\alpha}) b_{\o_{\alpha}}(p_{\beta}) \> \< d_{\o_{\alpha}}^{\dag}(p_{\beta}) d_{\o_{\alpha}}(p_{\alpha}) \> \Big\} \Big] N_{\o_{\g}}^{(1)}(p_{\gamma}),  \label{eq:I1} \nonumber \\
      &+& 2\delta_{ijk} \Big[ \Re \Big\{ \< b_{i}^{\dag}(p_1) b_{i}(p_2) \> \< b_{i}^{\dag}(p_2) b_{i}(p_3) \> \< b_{i}^{\dag}(p_3) b_{i}(p_1) \> \Big\}  + 3\Re \Big\{\< d_{i}^{\dag}(p_1) d_{i}(p_2) \> \< b_{i}^{\dag}(p_2) b_{i}(p_3) \> \< b_{i}^{\dag}(p_3) b_{i}(p_1) \> \Big\} \Big], \\
      I_2 &=& \sum_{\a} \< b_{\o_{\a}}^{\dag}(p_{\a}) b_{\o_{\a}}(p_{\a}) \> \Big[ \< \prod_{\epsilon \in \o \setminus \{\a\} } d_{\o_{\epsilon}}^{\dag}(p_{\epsilon}) d_{\o_{\epsilon}}(p_{\epsilon}) \> - \prod_{\epsilon \in \o \setminus \{\a\} } \< d_{\o_{\epsilon}}^{\dag}(p_{\epsilon}) d_{\o_{\epsilon}}(p_{\epsilon}) \> \Big], \nonumber \\
      &+& 2 \sum_{\o} \delta_{\o_{\alpha}\o_{\beta}} \Re \Big\{ \Big[ \< d_{\o_{\alpha}}^{\dag}(p_{\alpha}) d_{\o_{\g}}^{\dag}(p_{\g}) d_{\o_{\alpha}}(p_{\beta}) d_{\o_{\g}}(p_{\g}) \> - \< d_{\o_{\alpha}}^{\dag}(p_{\alpha}) d_{\o_{\alpha}}(p_{\beta}) \> \< d_{\o_{\g}}^{\dag}(p_{\gamma}) d_{\o_{\g}}(p_{\gamma}) \> \Big] \< b_{\o_{\a}}^{\dag}(p_{\beta}) b_{\o_{\a}}(p_{\alpha}) \> \Big\}, \label{eq:I2} \\
      I_3 &=&  \< \prod_{\epsilon \in \o} d_{\o_{\epsilon}}^{\dag}(p_{\epsilon}) d_{\o_{\epsilon}}(p_{\epsilon}) \> - \prod_{\epsilon \in \o} \< d_{\o_{\epsilon}}^{\dag}(p_{\epsilon}) d_{\o_{\epsilon}}(p_{\epsilon}) \>, \label{eq:I3} 
    \end{eqnarray}
  \end{widetext}}
Summations over the set $\o$ represent all possible combinations of n-tuples, the size of which is given by the number of elements in the adjacent $\delta$ functions.
For example, a summation with $\delta_{\o_{\a}\o_{\b}}$ represents a sum over all pair combinations in the set where $\o_{\a}=\o_{\b}$.  
A summation with $\delta_{\o_{\a}\o_{\b}} \delta_{\o_{\g}\o_{\t}}$ represents a sum over all double-pair combinations satisfying the respective pair constraint.
The set of isospin indices for the three- and four-pion distribution is denoted by $\o=\{i,j,k\}$ and $\o=\{i,j,k,l\}$, respectively.
The isospin indices represent one of three pion types: $\pi^+,\pi^-,\pi^0$.

The expressions for two-pion correlations were derived previously \cite{Biyajima:1998yh, Akkelin:2001nd}.
Also, the expression for the same-charge three-pion correlation function was derived in Ref.~\cite{Biyajima:1998yh}.
The expressions are appropriate for three- and four-pion correlation functions with any pion combination.  
One may also find a discussion of the $\pi^- \pi^- \pi^+ \pi^+$ QS correlation without coherent emission in Ref.~\cite{Lednicky:1982dz}.

For the rest of this letter we only consider charged pion correlation functions since the LHC measurement was done exclusively so \cite{Adam:2015pbc}.
The multipion phases \cite{Andreev:1992pu,Heinz:1997mr} are neglected as the measurement of the $r_3$ function, constructed to isolate the three-pion phase, did not indicate a significant decrease with $Q_3$ \cite{Abelev:2013pqa}.
There also exits a possible phase between the chaotic and coherent component which is neglected here.

The coherent fraction, $G$, is typically taken to be momentum dependent and is also indicated to be such by the ALICE measurement \cite{Adam:2015pbc}.
An important additional observation is the relative momentum dependence to $G$ \cite{pubnote:2015}.
The analysis indicates that $G$ decreases quite rapidly with increasing $Q_{3,4}$, indicating that an emission of coherent pions might be collimated.
Furthermore, it was noted that the $\left< \pT \right>$ changes very little with $Q_n$ for a fixed $K_{\rm Tn}$ interval of the analysis.
Being such, the usual momentum dependent coherent fraction, $G(p)$, is promoted to $G(Q_n)$, which is appropriate for correlation functions in sufficiently narrow $K_{\rm Tn}$ intervals. 
It is important to note that the possibility of multiple coherent sources \cite{Ikonen:2008zz} can be difficult to distinguish from a single coherent source in the presence of collimated emission.
The quantum interference between two independently coherent sources radiating back-to-back would be experimentally unobservable.

The functional form of the coherent fraction was parametrized adequately with a simple Gaussian form: $G(Q_n) = \alpha e^{-(\beta Q_n)^2}$ \cite{pubnote:2015}.
The coherent fractions were extracted from five types of same-charge correlation functions \cite{Gangadharan:2015ina}, $C_3^{\rm QS}, c_3^{\rm QS}, C_4^{\rm QS}, a_4^{\rm QS}$, and $b_4^{\rm QS}$, each of which yield different Gaussian parameters.  
Additionally, two extreme assumptions of the size of the coherent component were considered; one where $R_{\rm coh}=R_{\rm ch}$, and the other where $R_{\rm coh}=0$.
For each assumption of $R_{\rm coh}$, the Gaussian fits to $C_3^{\rm QS}$ and $c_3^{\rm QS}$ are used to form two alternate forms of $G(Q_3)$.
Likewise, the Gaussian fits to $C_4^{\rm QS}$ and $b_4^{\rm QS}$ are used to form two alternate forms of $G(Q_4)$.
The Gaussian parameters are shown in Tab.~\ref{tab:GaussG}.
\begin{table}
  \center
  \begin{tabular}{| c | c | c | c | c |}
    \hline
    $G$ from & $C_3^{\rm QS}$ & ${\bf c}_3^{\rm QS}$ & $C_4^{\rm QS}$ & $b_4^{\rm QS}$ \\ \hline
    $\alpha$ & $71 (74)$ & $49 (46)$ & $37 (29)$ & $59 (59)$  \\ \hline
    $\beta$ & $20 (27)$ & $16 (23)$ & $6 (11)$ & $20 (24)$ \\ \hline
  \end{tabular}
  \caption{Gaussian fit parameters of the coherent fraction versus relative momentum: $G(Q_n)=\alpha e^{-(\beta Q_n)^2}$ \cite{pubnote:2015}.  Parameters from the four listed  correlation functions are shown with two extreme assumptions of the coherent source radius: $R_{\rm coh}=R_{\rm ch}$ ($R_{\rm coh}=0$).}
\label{tab:GaussG}
\end{table}

Inserting Eqs.~\ref{eq:Tij} - \ref{eq:tijmixed} into Eqs.~\ref{eq:I1}-\ref{eq:I3} and Eqs.~\ref{eq:J1}-\ref{eq:J4} and with the values of the isospin vector averages in Eq.~\ref{eq:en}, we arrive at the three- and four-pion QS correlation functions,
{\footnotesize
\begin{widetext}
\begin{eqnarray}
  C_{2,ij}^{\rm QS}(\vecp_1,\vecp_2) &=& \frac{N_{ij}^{(2)}(\vecp_1,\vecp_2)}{N_i^{(1)}(\vecp_1)N_j^{(1)}(\vecp_2)} = 1 + \delta_{ij} \Big[ 2G(Q_2)[1-G(Q_2)]T_{12}t_{12} + [1-G(Q_2)]^2 T_{12}^2 \Big] + \xi_{22}, \label{eq:C2iso} \\
  C_{3,ijk}^{\rm QS}(\vecp_1,\vecp_2,\vecp_3) &=& \frac{I_1 + I_2 + I_3}{N_i^{(1)}(\vecp_1)N_j^{(1)}(\vecp_2)N_k^{(1)}(\vecp_3)},  \nonumber \\
  &=& 1 + [1-G(Q_3)] \sum_{\o} \delta_{\o_{\alpha}\o_{\beta}} \Big[ [1-G(Q_3)]T_{\alpha\beta}^2 + 2G(Q_3) T_{\alpha\beta}t_{\alpha\beta} \Big], \nonumber \\
  &+& 2 \delta_{ijk} [1-G(Q_3)]^2 \Big[ [1-G(Q_3)] T_{12}T_{23}T_{31} + 3G(Q_3)T_{12}T_{23}t_{31} \Big], \nonumber \\
  &+& \xi_{23} \Big\{ 3 + 2\sum_{\o} \delta_{\o_{\alpha}\o_{\beta}} T_{\alpha\beta}t_{\alpha\beta} \Big\} + \xi_{33},  \label{eq:C3iso} \\
  C_{4,ijkl}^{\rm QS}(\vecp_1,\vecp_2,\vecp_3,\vecp_4) &=& \frac{J_1 + J_2 + J_3 + J_4}{N_i^{(1)}(\vecp_1)N_j^{(1)}(\vecp_2)N_k^{(1)}(\vecp_3)N_l^{(1)}(\vecp_4)}, \nonumber \\
  &=& 1 + [1-G(Q_4)] \sum_{\o} \delta_{\o_{\alpha}\o_{\beta}} \Big[ [1-G(Q_4)]T_{\alpha\beta}^2 + 2G(Q_4) T_{\alpha\beta}t_{\alpha\beta} \Big], \nonumber \\
  &+& 2 \sum_{\o} \delta_{\o_{\a}\o_{\b}\o_{\g}} \Big[ [1-G(Q_4)]^3T_{\a\b}T_{\b\g}T_{\g\a} + 3G(Q_4)[1-G(Q_4)]^2t_{\a\b}T_{\b\g}T_{\g\a} \Big], \nonumber \\
  &+& \sum_{\o} \delta_{\o_{\a}\o_{\b}} \delta_{\o_{\g}\o_{\t}}  \Big[ [1-G(Q_4)]^4 T_{\a\b}^2 T_{\b\g}^2 + 2G(Q_4)[1-G(Q_4)]^3 (T_{\a\b}^2 T_{\g\t} t_{\g\t} + T_{\g\t}^2 T_{\a\b} t_{\a\b}),  \label{eq:sp1} \nonumber \\ 
  &+& 4 G^2(Q_4)[1-G(Q_4)]^2 T_{\a\b}t_{\a\b}T_{\g\t}t_{\g\t}  \Big],  \label{eq:sp2} \nonumber \\ 
  &+& 2 \delta_{\o_{\a}\o_{\b}\o_{\g}\o_{\t}} [1-G(Q_4)]^3\Big[ [1-G(Q_4)]T_{\a\b}T_{\b\g}T_{\g\t}T_{\t\a} + 4G(Q_4)t_{\a\b}T_{\b\g}T_{\g\t}T_{\t\a} + \mlq \b \rightleftharpoons \g \mrq + \mlq \g \rightleftharpoons \t \mrq \Big], \nonumber \\
  &+& \xi_{24} \Big\{ 6 + \sum_{\o} \delta_{\o_{\a}\o_{\b}}\Big[ T_{\a\b}^2 + 4T_{\a\b}t_{\a\b} + 4 \delta_{\o_{\g}\o_{\t}} T_{\a\b}t_{\a\b}T_{\g\t}t_{\g\t}  \Big], \label{eq:sp3} \nonumber \\ 
  &+& 2 \sum_{\o} \delta_{\o_{\a}\o_{\b}\o_{\g}} \Big[ T_{\a\b}T_{\b\g}t_{\g\a} + T_{\a\b}t_{\b\g}T_{\g\a} + t_{\a\b}T_{\b\g}T_{\g\a} \Big] \Big\}, \nonumber \\ 
  &+& \xi_{34} \Big\{4 + 2\sum_{\o} \delta_{\o_{\a}\o_{\b}}t_{\a\b}T_{\a\b} \Big\} + \xi_{44}. \label{eq:C4iso}
\end{eqnarray}
\end{widetext}}
The additional terms due to isospin conservation appear with the coefficient: $\xi_{mn}=\left[ 3^{m}\< |e_+|^{2m} \> - 1 \right] G^{m}(Q_n)[1-G(Q_n)]^{n-m}$.
The functional form of $G(Q_2)$ is left ambiguous here as it is was not directly measured.  
However, one may approximate it with $G(Q_{3,4})$ in the symmetric configuration of each pair $Q_2$.

\section{Mixed-charge cumulant correlation functions}
The standard $n$-pion correlation functions can not easily be used to extract the isospin conservation induced correlations as they also contain the full set of BE correlations.
The cumulant $n$-pion correlation functions, on the other hand, are defined such that all lower order ($<n$) symmetrization sequences not coupled to the $\xi_{mn}$ terms are explicitly removed \cite{Gangadharan:2015ina}. 
Mixed-charge cumulant correlation functions therefore present a unique advantage since the entire set of symmetrizations are of a lower order and removed in its construction; making isospin correlations easier to identify.

Using Eqs.~\ref{eq:C2iso} - \ref{eq:C4iso} and rearranging terms in powers of $G$, one obtains the following expressions for the three- and four-pion mixed-charge cumulant correlation functions,
{\footnotesize
\begin{widetext}
\begin{eqnarray}
c_{3,\pi^- \pi^- \pi^+}^{\rm QS} &=& \Big[ N_{\pi^- \pi^- \pi^+}^{(3)}(Q_3) - N_{\pi^- \pi^-}^{(2)}N_{\pi^+}^{(1)}(Q_3) - 2N_{\pi^- \pi^+}^{(2)}N_{\pi^-}^{(1)}(Q_3) +  3N_{\pi^-}^{(1)}N_{\pi^-}^{(1)}N_{\pi^+}^{(1)}(Q_3) \Big] / N_{\pi^-}^{(1)}N_{\pi^-}^{(1)}N_{\pi^+}^{(1)}(Q_3), \nonumber \\ 
  &=& 1 + \frac{2}{5}T_{12}t_{12}G^2(Q_3) - \Big[\frac{2}{35} + \frac{2}{5}T_{12}t_{12} \Big] G^3(Q_3), \label{eq:c3MC} \\
  c_{4,\pi^- \pi^- \pi^- \pi^+}^{\rm QS} &=& \Big[ N_{\pi^- \pi^- \pi^- \pi^+}^{(4)}(Q_4) - N_{\pi^- \pi^- \pi^-}^{(3)}N_{\pi^+}^{(1)}(Q_4) - 3N_{\pi^- \pi^+}^{(2)}N_{\pi^-}^{(1)}N_{\pi^-}^{(1)}(Q_4) + 4N_{\pi^-}^{(1)}N_{\pi^-}^{(1)}N_{\pi^-}^{(1)}N_{\pi^+}^{(1)}(Q_4) \Big] \nonumber \\
  &/& N_{\pi^-}^{(1)}N_{\pi^-}^{(1)}N_{\pi^-}^{(1)}N_{\pi^+}^{(1)}(Q_4), \nonumber \\
  &=& 1 + \frac{G^2(Q_4)}{5}\Big[ T_{12}^2 + 2T_{12}t_{12} + 2T_{12}T_{23}t_{13} \Big] - \frac{G^3(Q_4)}{35}\Big[ 6 + 14T_{12}^2  + 4T_{12}t_{12} + 28T_{12}T_{23}t_{13} \Big], \nonumber \\
  &+& \frac{G^4(Q_4)}{35}\Big[ 3 + 7T_{12}^2 - 10T_{12}t_{12} + 14T_{12}T_{23}t_{13} \Big] + \mlq 12 \rightleftharpoons 13 \mrq + \mlq 12 \rightleftharpoons 23 \mrq ,\label{eq:c4MC1} \\
  c_{4,\pi^- \pi^- \pi^+ \pi^+}^{\rm QS} &=& \Big[ N_{\pi^- \pi^- \pi^+ \pi^+}^{(4)}(Q_4) - N_{\pi^- \pi^-}^{(2)}N_{\pi^+ \pi^+}^{(2)}(Q_4) - 4N_{\pi^- \pi^+}^{(2)}N_{\pi^-}^{(1)}N_{\pi^+}^{(1)}(Q_4) + 5N_{\pi^-}^{(1)}N_{\pi^-}^{(1)}N_{\pi^+}^{(1)}N_{\pi^+}^{(1)}(Q_4) \Big] \nonumber \\
  &/& N_{\pi^-}^{(1)}N_{\pi^-}^{(1)}N_{\pi^+}^{(1)}N_{\pi^+}^{(1)}(Q_4), \label{eq:c4MC2pre} \nonumber \\
  &=& 1 + \frac{G^2(Q_4)}{5}\Big[4T_{12}t_{12} + 4T_{34}t_{34} + 4T_{12}t_{12}T_{34}t_{34}  \Big] - \frac{G^3(Q_4)}{35}\Big[ 8 + 32(T_{12}t_{12} + T_{34}t_{34}) + 56T_{12}t_{12}T_{34}t_{34}  \Big], \nonumber \\
  &+& \frac{G^4(Q_4)}{175}\Big[ 8 + 20(T_{12}t_{12} + T_{34}t_{34}) + 140 T_{12}t_{12}T_{34}t_{34}  \Big]. \label{eq:c4MC2}
\end{eqnarray}
\end{widetext}}
Cyclic permutations of all pair-exchange magnitudes are indicated by $\mlq 12 \rightleftharpoons 13 \mrq$ and $\mlq 12 \rightleftharpoons 23 \mrq$.
The relevant normalizations to make the calculations comparable to the experimental procedure can be found from Eqs.~\ref{eq:c3MC} - \ref{eq:c4MC2} by setting $T_{ij},t_{ij}$, and $G(Q_n)$ to their respective values at large relative momentum.  
In the ALICE analysis, the normalization region for central Pb--Pb collisions is given by $Q_{2,ij}=0.175$ MeV/$c$.
Concerning the term $N_{\pi^- \pi^-}^{(2)}N_{\pi^+ \pi^+}^{(2)}(Q_4)$, the coherent fraction is parameterized as $G(Q_4)$, not as $G(Q_2)$.

\section{Results} \label{sec:Results}
\subsection{$\pi^- \pi^+$ correlations}
Before presenting multipion isospin conservation induced correlations, two-pion correlations need to be discussed, especially in regards to the application of FSI corrections in the ALICE analysis.
In the presence of coherent pion emission, it is clear that $\pi^+ \pi^-$ correlations contain positive contributions from not only FSI, but isospin correlations (Eq.~\ref{eq:C2iso}) as well.
However, the correction for FSI in the ALICE analysis was performed assuming only FSI contributions to $\pi^+ \pi^-$ correlations.  
The standard form of the measured two-pion correlation function, $C_2$, can be written in terms of the QS+FSI correlation, $C_2^{\rm QS}$, as $C_2 = (1-f_c^2) + f_c^2 K_2 C_2^{\rm QS}$.
The parameter $f_c^2$ describes the correlated fraction of pairs.
As it was done, the FSI factor, $K_2$, was calculated within the \textsc{therminator} model \cite{Kisiel:2005hn,Chojnacki:2011hb} of particle freeze-out and only utilizing pion pairs with sufficiently small separation ($r^*<100$ fm) for which observable FSI correlations are expected.  
The remaining pairs with large separations were taken into account by tuning the $f_c$ parameter such that the FSI corrected $\pi^+ \pi^-$ correlation function was consistent with unity.
Such a procedure would clearly result in an {\it overcorrection} of FSI correlations in the presence of coherent pion emission.

\begin{figure}[!h]
\center
  \includegraphics[width=0.44\textwidth]{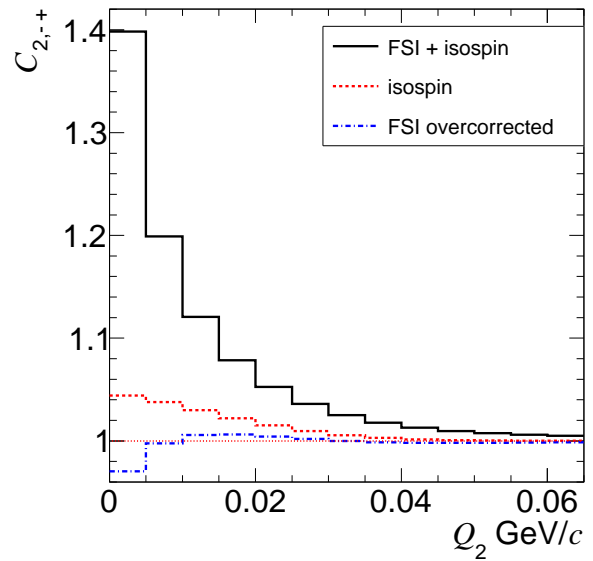}
  \caption{$\pi^- \pi^+$ correlations.  The solid black line shows the full correlation (FSI + isospin).  The dashed red line shows the FSI corrected correlation while the dot-dashed blue line shows the FSI overcorrected correlation.}
  \label{fig:C2MC}
\end{figure}

To demonstrate this scenario, we model the coherent fraction in a $Q_2$ dependent form using the parameters $\alpha=0.49, \beta=16$ from Tab.~\ref{tab:GaussG}. 
Although these parameters were extracted from the measured three-pion correlations, it is assumed that a similar form exists for $G(Q_2)$ and can be approximated by $G(Q_2)=\alpha e^{-(Q_3 \beta)/\sqrt{3}}$.
Figure \ref{fig:C2MC} shows the FSI+isospin, pure isospin, and an FSI overcorrected $\pi^+ \pi^-$ correlation function.  
The FSI factor was enhanced by $30\%$ ($(K_2-1)\times1.3 + 1$) to obtain the overcorrected correlation.
An overcorrection will also distort the extracted multipion correlations.
The multipion FSI factor in the ALICE analysis was treated as the product of pair FSI factors, $K_3=K_2^{12} K_2^{13} K_2^{23}$, $K_4=K_2^{12} K_2^{13} K_2^{14}K_2^{23} K_2^{24} K_2^{34}$.
Here we use the same product of pairs treatment of multipion FSI which can also be factored from the QS correlations.

\subsection{Mixed-charge multipion correlations}
Calculations of the mixed-charge cumulant correlations are now presented and compared to the ALICE data.
Three ingredients are needed for the calculation: $G(Q_n)$, $T_{ij}$, and $t_{ij}$.  
The Gaussian parametrization of $G(Q_n)$ \cite{pubnote:2015}, was already discussed.
An Edgeworth expansion \cite{Csorgo:2000pf} is used to parametrize $T_{ij}$:  
\begin{eqnarray}
T_{ij} &=& sE_{\rm w}(R_{\rm ch}\,Q_{2,ij})\,e^{-R_{\rm ch}^2\,Q_{2,ij}^2/2} \label{eq:EWEA} \\
E_{\rm w}(R_{\rm ch},Q_{2,ij}) &=& 1 + \sum_{n=3}^{\infty} \frac{\kappa_n}{n! (\sqrt{2})^n} H_n(R_{\rm ch}\,Q_{2,ij}),
\end{eqnarray}
where $H_n$ are the Hermite polynomials, and $\kappa_n$ are the Edgeworth coefficients.  
The values of the parameters suggested by the most central ALICE data \cite{Abelev:2013pqa} for $0.2<\KTTwo<0.3$ GeV/$c$ are: $R_{\rm ch}=10.5$ fm, $\kappa_3=0.14$, $\kappa_4=0.29$.  
To ensure that $T_{ij}(Q_2=0)=1$, the $s$ parameter is normalized such that $s=1/(1 + \kappa_4/8)$.
The pair-exchange magnitude of the coherent component, $t_{ij}$, is considered in two extreme cases.  
For the case when $R_{\rm coh}=R_{\rm ch}$, $t_{ij}=T_{ij}$.  
For the case when $R_{\rm coh}=0$, $t_{ij}=1$.

In the experimental data, the effects of track-merging and splitting are minimized by removing pairs of particles with an angular separation below a certain threshold\cite{Aamodt:2011mr, Adam:2015pbc}.  
The cut is not applied to mixed-charge pairs as the effect is negligible for charged pions which curve in opposite directions in a solenoidal magnetic field.
The resulting imbalance of same- and mixed-charge pairs in the mixed-charge multipion correlation functions is reproduced here with the same pair cuts.

\begin{figure}[!h]
\center
  \includegraphics[width=0.44\textwidth]{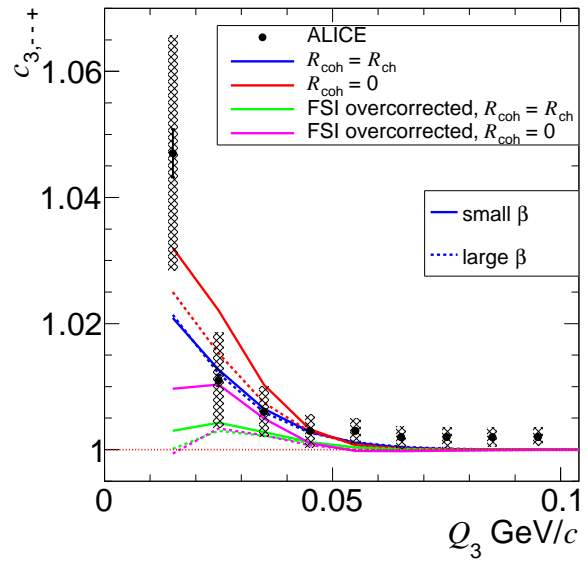}
  \caption{Mixed-charge three-pion cumulant correlations in ALICE compared to the isospin calculation.  The ALICE data is shown with black points for which the gray bands correspond to the systematic uncertainties.  Blue lines correspond to the isospin calculation where $R_{\rm coh}=R_{\rm ch}$ and red lines correspond to $R_{\rm coh}=0$.  Green and magenta lines represent the FSI overcorrected version.  Solid lines correspond to the smaller $\beta$ settings in Tab.~\ref{tab:GaussG}.}
  \label{fig:c3MC}
\end{figure}

The mixed-charge three-pion cumulant correlations in ALICE are compared to the isospin calculations in Figs.~\ref{fig:c3MC}.
Blue and red lines present the true isospin calculation while the green and magenta lines show an FSI overcorrected version, which were obtained by enhancing $K_2$ by $10\%$ unlike the $30\%$ used in Fig.~\ref{fig:C2MC} for reasons to be explained later. 
Given the large systematic uncertainties of the data, one cannot determine which parameterization of the coherent component is preferred. 

\begin{figure}[!h]
  \centering
  \subfigure[ $\pi^- \pi^- \pi^- \pi^+$]{
    \includegraphics[width=0.44\textwidth]{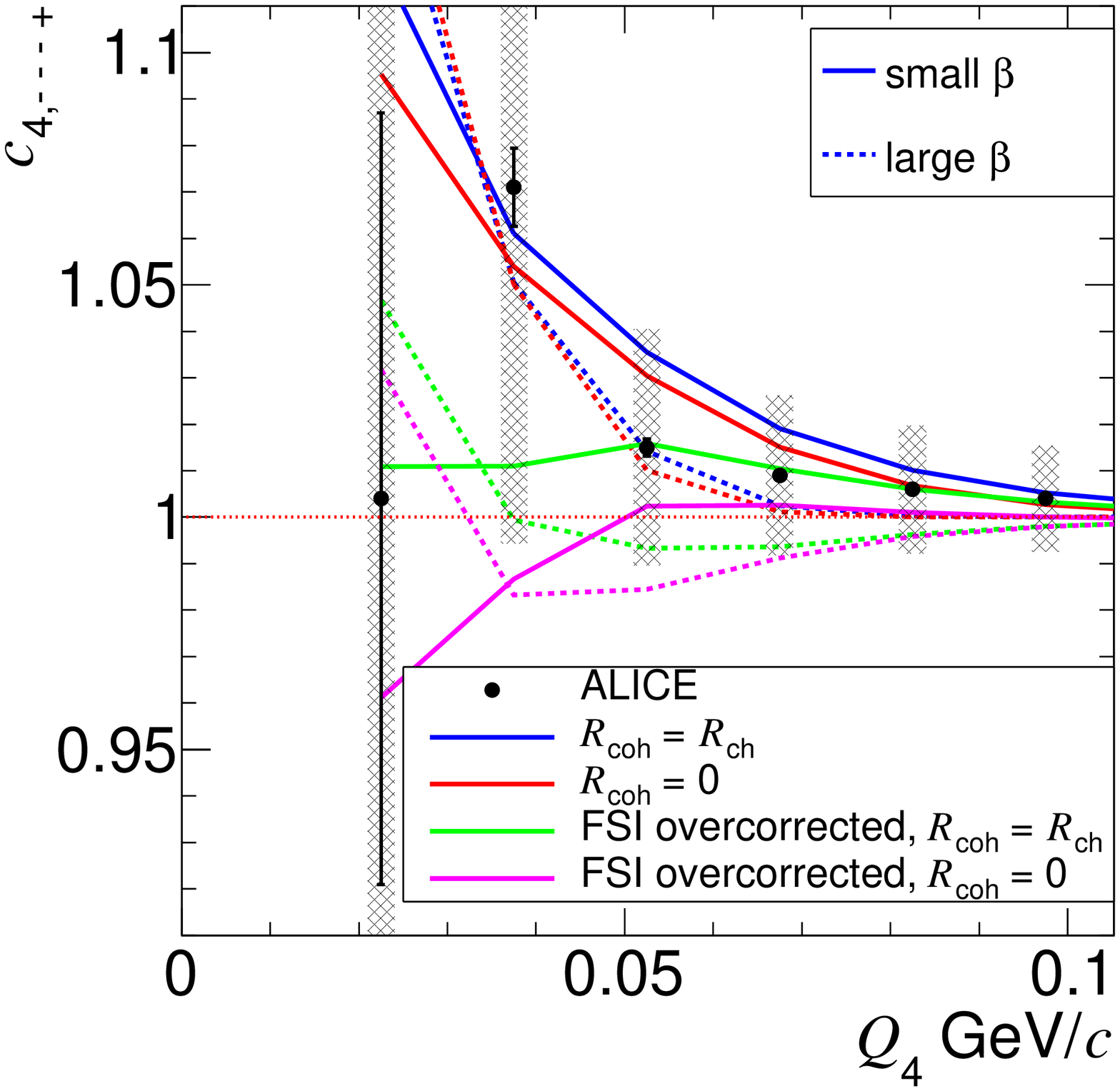}
    \label{fig:c4MC1}
  }
  \subfigure[ $\pi^- \pi^- \pi^+ \pi^+$]{
    \includegraphics[width=0.44\textwidth]{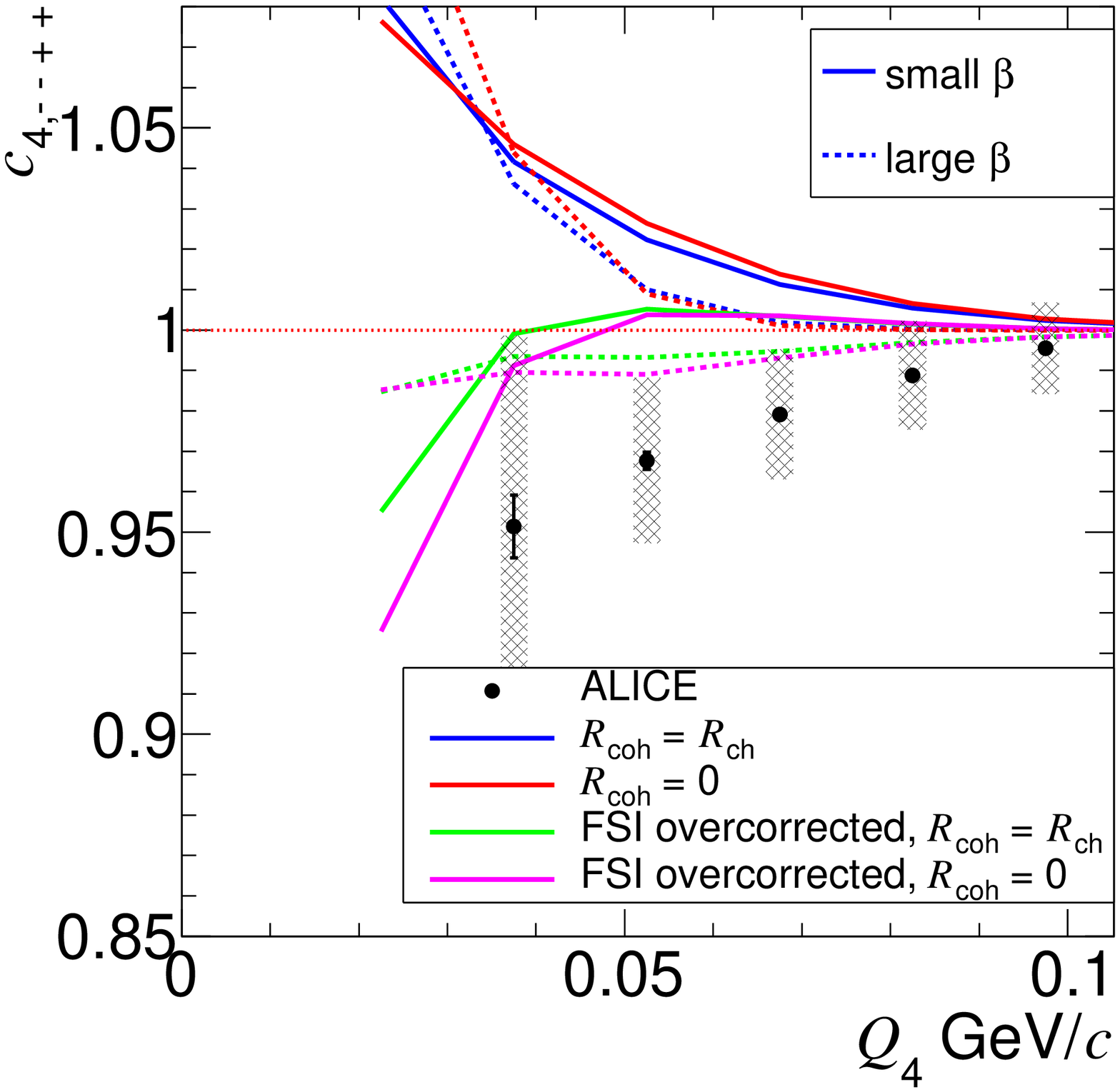}
    \label{fig:c4MC2}
  }
\caption{Mixed-charge four-pion cumulant correlations in ALICE compared to the isospin calculation.  The lowest $Q_4$ ALICE data point in Fig.~\ref{fig:c4MC2} is off-scale at 1.3 with very large systematic uncertainties.  The other details are the same as in Fig.~\ref{fig:c3MC}.}
\end{figure}

The mixed-charge four-pion cumulant correlations in ALICE are compared to the isospin calculations in Figs.~\ref{fig:c4MC1} - \ref{fig:c4MC2}.
In Fig.~\ref{fig:c4MC1}, the ALICE data are consistent with each curve due to the large systematic uncertainties.
$K_2$ was enhanced by $10\%$ to obtain the overcorrection.  Increasing the overcorrection even more would drive the calculations more negative and further away from the ALICE data. 
However, for $\pi^- \pi^- \pi^+ \pi^+$ correlations in Fig.~\ref{fig:c4MC2}, the data seem to agree best with the FSI overcorrected version.

The current ambiguity of the pair FSI factor, $K_2$, as well as the $f_c$ parameter obstruct a definitive statement of the observation of isospin conservation induced correlations from coherent emission.
Given the negative values of the measured $\pi^- \pi^- \pi^+ \pi^+$ cumulant correlations, an overcorrection of FSI correlations is indeed plausible.  
In order for the experimental data to better match the calculations, $f_c$ would have to be increased and $K_2$ decreased beyond the interval considered in the ALICE analysis.
The default setting for $f_c$ was 0.837 with a variation of 0.03.  
It is estimated that increasing $f_c$ by $10\%$ while simultaneously decreasing $K_2$ by $30\%$ would bring the extracted experimental mixed-charge correlations in line with the isospin calculations.

The isospin effect for same-charge correlations has also been calculated in the context of ``measured" versus ``built" correlations \cite{Gangadharan:2015ina}.
Two-pion correlations were first computed according to Eq.~\ref{eq:C2iso} and then used to build multipion QS correlations without the isospin effect as it was done in Ref.~\cite{Adam:2015pbc}.
The resulting bias depends on relative momentum as well as the coherent fraction.  
For the lowest relative momentum intervals in the ALICE analysis, and with the extracted coherent fractions, a bias of less than $2\%$ is expected.

\section{Summary}
The suppression of multipion Bose-Einstein correlations at the LHC may indicate a fundamentally new property of heavy-ion collisions.
The possibility of quantum coherence to explain the suppression has been considered here in regards to the complimentary feature of isospin conservation of coherent pion emission.
Given the coherent fractions extracted from the suppression of same-charge multipion measurements, the isospin correlations for mixed-charge cumulants have been calculated.

The $\pi^- \pi^- \pi^+$ and $\pi^- \pi^- \pi^- \pi^+$ calculations generally agree with the ALICE data although the large experimental uncertainties prevent a definitive statement.  
For the $\pi^- \pi^- \pi^+ \pi^+$ case, the experimental correlations are negative which can be explained by an FSI overcorrection of the data. 
Such an overcorrection may be expected since the pair FSI correction, which are used to construct the multipion FSI factors, were tuned such that the $\pi^+ \pi^-$ correlation function was consistent with unity after FSI corrections.
However, isospin correlations are also expected for $\pi^+ \pi^-$ correlations, thus necessitating a positive residue after FSI corrections.
In order to bring the experimentally extracted $\pi^+ \pi^-$, $\pi^- \pi^- \pi^+$, $\pi^- \pi^- \pi^- \pi^+$, and $\pi^- \pi^- \pi^+ \pi^+$ correlations inline with the true isospin correlations, the two factors which control the FSI correction, $f_c$ and $K_2$, would need to be altered beyond the experimentally considered interval.
It is estimated that $f_c$ would need to increase by about $10\%$ and the FSI cumulant $(K_2-1)$ decrease by $30\%$.
The alteration effectively shifts the treatment of long-lived emitters from $f_c$ to $K_2$.

Despite evidence for both the suppression of Bose-Einstein correlations as well as the complimentary isospin correlations, coherent pion emission poses several conceptual difficulties. 
Hydrodynamic models, which assume local thermal equilibrium, have been successful in describing a wide variety of measurements in heavy-ion collisions. 
However, the scattering needed to equilibrate the medium and produce the collective expansion is also expected to destroy coherence.
In addition, the exact mechanism to generate large coherent fractions at low relative momentum remains unknown.

\newenvironment{acknowledgement}{\relax}{\relax}
\begin{acknowledgement}
\section*{Acknowledgments}
This material is based upon work supported in part by the U.S. Department of Energy, Office of Science, Office of Nuclear Physics, under contract number DE-AC02-05CH11231.
\end{acknowledgement}

\section{Appendix}
\label{app:supp}
Each of the four terms given in Eq.~\ref{eq:J1234} are given below.
The nomenclature is the same as that used in Eqs.~\ref{eq:I1}-\ref{eq:I3}.
{\scriptsize
  \begin{widetext}
    \begin{eqnarray}
           J_1 &=&  N_i^{(1)}(p_1)N_j^{(1)}(p_2)N_k^{(1)}(p_3)N_l^{(1)}(p_4),  \label{eq:J1} \\
      &+& \sum_{\o} \delta_{\o_{\alpha}\o_{\beta}} \Big[ |\< b_{\o_{\alpha}}^{\dag}(p_\alpha) b_{\o_{\alpha}}(p_\beta) \>|^2 + 2\Re \Big\{ \< b_{\o_{\alpha}}^{\dag}(p_\beta) b_{\o_{\alpha}}(p_\alpha) \> \< d_{\o_{\alpha}}^{\dag}(p_\beta) d_{\o_{\alpha}}(p_\alpha) \> \Big\} \Big] \prod_{\epsilon \notin \{\a, \b \} } N_{\o_{\epsilon}}^{(1)}(p_\epsilon),  \nonumber \\
      &+& 2\sum_{\o} \delta_{\o_{\alpha}\o_{\beta}\o_{\gamma}} \Big[ \Re \Big\{ \< b_{\o_{\alpha}}^{\dag}(p_\alpha) b_{\o_{\alpha}}(p_\beta) \> \< b_{\o_{\alpha}}^{\dag}(p_\beta) b_{\o_{\alpha}}(p_\gamma) \> \< b_{\o_{\alpha}}^{\dag}(p_\gamma) b_{\o_{\alpha}}(p_\alpha) \> \Big\},  \nonumber \\
      &+& 3\Re \Big\{ \< d_{\o_{\alpha}}^{\dag}(p_\alpha) d_{\o_{\alpha}}(p_\beta) \> \< b_{\o_{\alpha}}^{\dag}(p_\beta) b_{\o_{\alpha}}(p_\gamma) \> \< b_{\o_{\alpha}}^{\dag}(p_\gamma) b_{\o_{\alpha}}(p_\alpha) \> \Big\} \Big] N_{\o_{\tau}}^{(1)}(p_\tau), \nonumber \\
      &+& \sum_{\o} \delta_{\o_{\alpha}\o_{\beta}} \delta_{\o_{\gamma}\o_{\tau}} \Big[ |\< b_{\o_{\alpha}}^{\dag}(p_{\alpha}) b_{\o_{\alpha}}(p_{\beta}) \>|^2 |\< b_{\o_{\gamma}}^{\dag}(p_{\gamma}) b_{\o_{\gamma}}(p_{\tau}) \>|^2 + 2|\< b_{\o_{\alpha}}^{\dag}(p_{\alpha}) b_{\o_{\alpha}}(p_{\beta}) \>|^2 \Re \Big\{ \< b_{\o_{\g}}^{\dag}(p_{\g}) b_{\o_{\g}}(p_{\t}) \> \< d_{\o_{\g}}^{\dag}(p_{\t}) d_{\o_{\g}}(p_{\g}) \> \Big\} ,  \label{eq:DP1} \nonumber \\
      &+& 2|\< b_{\o_{\g}}^{\dag}(p_{\g}) b_{\o_{\g}}(p_{\t}) \>|^2 \Re \Big\{ \< b_{\o_{\a}}^{\dag}(p_{\a}) b_{\o_{\a}}(p_{\b}) \> \< d_{\o_{\a}}^{\dag}(p_{\b}) d_{\o_{\a}}(p_{\a}) \> \Big\} ,  \label{eq:DP2} \nonumber \\
        &+& 4 \Re \Big\{ \< b_{\o_{\alpha}}^{\dag}(p_{\a}) b_{\o_{\alpha}}(p_{\b}) \> \< d_{\o_{\a}}^{\dag}(p_{\b}) d_{\o_{\a}}(p_{\a}) \> \Big\} \Re \Big\{ \< b_{\o_{\g}}^{\dag}(p_{\g}) b_{\o_{\g}}(p_{\t}) \> \< d_{\o_{\g}}^{\dag}(p_{\t}) d_{\o_{\g}}(p_{\g}) \> \Big\}  \Big], \label{eq:DP3} \nonumber \\ 
      &+&  2\delta_{\o_{\alpha}\o_{\beta}\o_{\gamma}\o_{\tau}} \Big[ \Re\Big\{ \< b_{\o_{\alpha}}^{\dag}(p_\alpha) b_{\o_{\alpha}}(p_\beta) \> \< b_{\o_{\alpha}}^{\dag}(p_\beta) b_{\o_{\alpha}}(p_\gamma) \> \< b_{\o_{\alpha}}^{\dag}(p_\gamma) b_{\o_{\alpha}}(p_\tau) \> \> \< b_{\o_{\alpha}}^{\dag}(p_\tau) b_{\o_{\alpha}}(p_\alpha) \> \Big\},  \nonumber \\
      &+& 4\Re \Big\{ \< d_{\o_{\alpha}}^{\dag}(p_\alpha) d_{\o_{\alpha}}(p_\beta) \> \< b_{\o_{\alpha}}^{\dag}(p_\beta) b_{\o_{\alpha}}(p_\gamma) \> \< b_{\o_{\alpha}}^{\dag}(p_\gamma) b_{\o_{\alpha}}(p_\tau) \> \< b_{\o_{\alpha}}^{\dag}(p_\tau) b_{\o_{\alpha}}(p_\alpha) \> \Big\} + \mlq \beta \rightleftharpoons \g \mrq + \mlq \g \rightleftharpoons \tau \mrq \Big], \nonumber \\
      J_2 &=& \Big( \< \prod_{\e \in \{ \a,\b \}} d_{\o_{\epsilon}}^{\dag}(p_{\epsilon}) d_{\o_{\epsilon}}(p_{\epsilon}) \> - \prod_{\epsilon \in \{\a,\b \}} \< d_{\o_{\epsilon}}^{\dag}(p_{\epsilon}) d_{\o_{\epsilon}}(p_{\epsilon}) \> \Big)  \Big( \prod_{\e \in \o \setminus \{\a,\b \}} \< b_{\o_{\epsilon}}^{\dag}(p_{\epsilon}) b_{\o_{\epsilon}}(p_{\epsilon}) \> + \Big[ \sum_{ \o \setminus \{\a,\b \} } \delta_{\o_{\g} \o_{\t}} | \< b_{\o_{\g}}^{\dag}(p_{\g}) b_{\o_{\g}}(p_{\t}) \> |^2 \Big]  \Big), \\
      &+& 4 \sum_{\o} \delta_{\o_{\a}\o_{\b}} \Re \Big\{ \Big[ \< d_{\o_{\a}}^{\dag}(p_{\a}) d_{\o_{\g}}^{\dag}(p_{\g}) d_{\o_{\a}}(p_{\b}) d_{\o_{\g}}(p_{\g}) \> - \< d_{\o_{\a}}^{\dag}(p_{\a}) d_{\o_{\a}}(p_{\b}) \> \< d_{\o_{\g}}^{\dag}(p_{\g}) d_{\o_{\g}}(p_{\g}) \> \Big] \< b_{\o_{\a}}^{\dag}(p_{\b}) b_{\o_{\a}}(p_{\a}) \> \< b_{\o_{\t}}^{\dag}(p_{\t}) b_{\o_{\t}}(p_{\t}) \> \Big\}, \nonumber \\
       &+& 4 \sum_{\o} \delta_{\o_{\a}\o_{\b}} \delta_{\o_{\g}\o_{\t}}   \nonumber \\ 
       &\times& \Re \Big\{ \Big[ \< d_{\o_{\a}}^{\dag}(p_{\a}) d_{\o_{\g}}^{\dag}(p_{\g}) d_{\o_{\a}}(p_{\b}) d_{\o_{\g}}(p_{\t}) \> - \< d_{\o_{\a}}^{\dag}(p_{\a}) d_{\o_{\a}}(p_{\b}) \> \< d_{\o_{\g}}^{\dag}(p_{\g}) d_{\o_{\g}}(p_{\t}) \> \Big] \< b_{\o_{\a}}^{\dag}(p_{\b}) b_{\o_{\a}}(p_{\a}) \> \< b_{\o_{\g}}^{\dag}(p_{\t}) b_{\o_{\g}}(p_{\g}) \> \Big\} , \label{eq:DP4} \nonumber \\ 	
      &+& 2\sum_{\o} \delta_{\o_{\a}\o_{\b}\o_{\g}} \Re \Big\{ \Big[ \< d_{\o_{\a}}^{\dag}(p_{\a}) d_{\o_{\t}}^{\dag}(p_{\t}) d_{\o_{\a}}(p_{\b}) d_{\o_{\t}}(p_{\t}) \> - \< d_{\o_{\a}}^{\dag}(p_{\a}) d_{\o_{\a}}(p_{\b}) \> \< d_{\o_{\t}}^{\dag}(p_{\t}) d_{\o_{\t}}(p_{\t}) \> \Big] \< b_{\o_{\a}}^{\dag}(p_{\b}) b_{\o_{\a}}(p_{\g}) \> \< b_{\o_{\a}}^{\dag}(p_{\g}) b_{\o_{\a}}(p_{\a}) \> \Big\}, \nonumber \\
      J_3 &=& \sum_{\a} \< b_{\o_{\a}}^{\dag}(p_{\a}) b_{\o_{\a}}(p_{\a}) \> \Big[ \< \prod_{\epsilon \in \o \setminus \{\a\} } d_{\o_{\epsilon}}^{\dag}(p_{\epsilon}) d_{\o_{\epsilon}}(p_{\epsilon}) \> - \prod_{\epsilon \in \o \setminus \{\a\} } \< d_{\o_{\epsilon}}^{\dag}(p_{\epsilon}) d_{\o_{\epsilon}}(p_{\epsilon}) \> \Big], \\
      &+& 2 \sum_{\o} \delta_{\o_{\a}\o_{\b}}  \Re\Big\{ \Big[ \< d_{\o_{\a}}^{\dag}(p_{\a}) d_{\o_{\a}}(p_{\b}) \prod_{\e \notin \{\a,\b\} } d_{\o_{\epsilon}}^{\dag}(p_{\epsilon}) d_{\o_{\epsilon}}(p_{\epsilon}) \> - \< d_{\o_{\a}}^{\dag}(p_{\a}) d_{\o_{\a}}(p_{\b}) \> \prod_{\e \notin \{\a,\b \}} \< d_{\o_{\epsilon}}^{\dag}(p_{\epsilon}) d_{\o_{\epsilon}}(p_{\epsilon}) \> \Big] \< b_{\o_{\a}}^{\dag}(p_{\b}) b_{\o_{\a}}(p_{\a}) \> \Big\}, \nonumber \\
      J_4 &=&  \< \prod_{\e \in \o} d_{\o_{\epsilon}}^{\dag}(p_{\epsilon}) d_{\o_{\epsilon}}(p_{\epsilon}) \> - \prod_{\e \in \o} \< d_{\o_{\epsilon}}^{\dag}(p_{\epsilon}) d_{\o_{\epsilon}}(p_{\epsilon}) \>. \label{eq:J4}
    \end{eqnarray}
  \end{widetext}}

\bibliography{biblio}{}

\end{document}